\documentclass[preprint,showpacs,preprintnumbers,amsmath,amssymb]{revtex4}

\usepackage{graphicx}
\usepackage{dcolumn}
\usepackage{bm}

\begin{document}

\title{Diffusion of small two-dimensional Cu Islands on Cu(111)}

\author{Altaf Karim, Ahlam N. Al-Rawi, Abdelkader Kara, Talat S. Rahman}
\affiliation{ Department of Physics, Cardwell Hall, Kansas State University,
Manhattan, Kansas 66506}
\author{Oleg Trushin}
\affiliation{Institute of Microelectronics and Informatics,
Academy of Sciences of Russia, Yaroslavl 150007, Russia }
\author{Tapio Ala-Nissila}
\affiliation{Laboratory of Physics, P.O. Box 1100, Helsinki
University of Technology, FIN-02015 TKK, Espoo, Finland, and
Department of Physics, Brown University, Providence R.I.
02912-1843}

\draft

\begin{abstract}
Diffusion of small two dimensional Cu islands (containing up to
$10$ atoms) on Cu(111) has been studied using the newly developed
self-learning Kinetic Monte Carlo ({\it SLKMC}) method. It is
based on a database of diffusion processes and their energetics
accumulated automatically during the implementation of the {\it
SLKMC} code. Results obtained from simulations in which atoms hop
from one fcc hollow site to another are compared with those
obtained from a parallel set of simulations in which the database
is supplemented by processes revealed in complementary molecular
dynamics simulations at $500$ K. They include processes involving
the hcp (stacking-fault) sites, which facilitate concerted motion
of the islands. A significant difference in the scaling of the
effective diffusion barriers with island size is observed between
the two cases. In particular, the presence of concerted island
motion leads to an almost linear increase in the effective
diffusion barrier with size, while its absence accounts for strong
size-dependent oscillations and anomalous behavior for trimers and
heptamers.
We also identify and discuss in detail the key microscopic
processes responsible for the diffusion and examine the
frequencies of their occurrence, as a function of island size and
substrate temperature.

\end{abstract}


\maketitle

\section{INTRODUCTION}

Acquiring a precise knowledge of the microscopic mechanisms
responsible for island diffusion or mass transport on surfaces is
an important step towards the understanding of phenomena such as
thin film growth and its morphological evolution. Motivated by
experimental observations, initially using field ion microscopy
(FIM) \cite{bas76} - \cite{kel93}, and more recently with the use
of the scanning tunneling microscope (STM) \cite{wen94}-
\cite{bus03}, the study of adatom and vacancy island diffusion as
a function of size has been an important concern also for many
theorists \cite{mue05} - \cite{salo}. Because of the inherent
differences in the microscopic processes responsible for the
diffusion and its scaling behavior with size, the discussion has
naturally bifurcated into those for the larger islands, usually
containing more than $20$ atoms, and the smaller ones ($N < 20$).
For the larger islands, the diffusion coefficients appear to scale
as a function of the size and the scaling exponent is expected to
reflect the intervening atomistic processes responsible for the
diffusion \cite{bog98,altaf}.  However, for the smaller islands a
consistent knowledge of the variation of their mobility with size
and the details of the responsible atomistic processes has not yet
been fully established, especially on the (111) surfaces of fcc
metals.

One of the distinguishing geometrical features of the fcc(111)
surface is the presence of two types of hollow sites: the so
called fcc site under which there is no atom in the second layer
and whose occupancy by an adatom maintains the crystal stacking
order, and the hcp site under which there is an atom in the second
layer and its nucleation can lead to a stacking fault. Whether or
not an adatom or atoms in an adatom island occupy one or the other
of these two sites depends on their relative occupation energies
and has significant consequences for epitaxial growth and the
morphological evolution of the surface.  Although Ag, Cu, Pt, and
Ir are all fcc crystals there is no guarantee that the adatom
would prefer to sit in the fcc site.  In fact, experiments show
that the hcp site is preferred on Ir(111) \cite{wan90,bus03},
while on Cu(111) the fcc site appears to be favored \cite{repp},
although the small difference in the occupation energy (a few meV)
between the two sites does not rule out the occupation of the hcp
sites.  For dimers, trimers, and other larger islands, mixed
occupancy of the two sites is also possible. The relative
probability of occupancy of these two sites on fcc(111) surfaces
continues to be the subject of much discussion and debate.

For small adatom islands earlier experimental studies point to a
general decrease in mobility with increasing island size, except
for some cases of anomalously large mobility \cite{bas76} -
\cite{kel93}. For larger islands short-range diffusion of the
atoms around the periphery, followed by adjustment of island
shape, has been proposed to be the dominant mechanism for the
diffusion \cite{wan90,liu92}. In the case of small Ir islands on
Ir(111), concerted gliding motion of the island has also been
reported \cite{wan90}.  Subsequent molecular dynamics (MD)
simulations using many-body potentials based on the embedded atom
method (EAM) \cite{foi86} have further disclosed that in addition
to gliding, there is simultaneous motion of a portion of the
island from the fcc to the hcp sites, creating a stacking fault
\cite{ham95}. The motion of the island could result from the rest
of the atoms shifting also to the available hcp site. On Ni(111),
for example, the smaller islands reportedly find the fcc to hcp
transition to be critical to their diffusion, although gliding of
the island as a whole and periphery motion has also been seen
\cite{ham95}.  In agreement with the experimental results of Wang
and Ehrlich on the higher mobility of tetramers as compared to
trimers for Ir on Ir(111), Chang {\it et al.} \cite{cha00} find
the barrier for diffusion for the tetramer to be lower than that
for the trimer for a number of fcc metals.  They also predict a
zigzag motion to be the dominant one for the dimer and the
tetramer, while predicting a concerted motion as a whole for the
trimer. Recent theoretical studies of the energetics and dynamics
of $1 - 7$ atom Cu islands on Cu(111) have once again highlighted
the role of the concerted motion of the island in controlling its
diffusion characteristics \cite{mar04}. In the very recent work of
Mueller {\it et al.} \cite{mue05} good agreement with experimental
data on submonolayer epitaxy on Ir(111) is also obtained with the
inclusion of concerted motion of islands (with the stacking fault
sites). Issues about the relative importance of the proposed
diffusion mechanisms, the relevance of the occupation of the hcp
sites, and the observed anomalous diffusion for certain sizes, are
striking aspects of the diffusion of small 2D islands on fcc(111)
surfaces and may control the subsequent growth patterns on these
surfaces.

Our purpose here is to determine the microscopic factors that
control the diffusion of small Cu islands on Cu(111) in an
unbiased manner. The focus in the work of Marinica {\it et al.}
\cite{mar04} is on application of a newer version of the EAM
potentials \cite{mis01} to calculate diffusion barriers and
pre-exponential factors for a set of likely processes that they
find from their MD simulations. However, Ref. \cite{mar04} ignores
the presence of mechanisms associated with atom-by-atom motion in
small islands and only considers a limited number of collective
processes to be responsible for diffusion. The diffusion
coefficient is thus simply obtained from application of the
Arrhenius law to the activation energy barrier and the diffusion
prefactor calculated for the chosen diffusion process. The natural
question is whether such {\it a priori} selection of the
responsible process precludes the contributions of other processes
and whether such exclusion makes any difference in the predicted
trends in island diffusion.  The deeper question is, of course,
whether it is possible to allow adatom islands to evolve as a
function of time with mechanisms of their own choosing and thereby
provide an unbiased illustration of the rate limiting step in the
diffusion and the relative contributions of various mechanisms. We
have recently developed a self-learning approach to kinetic Monte
Carlo (KMC) simulations {\it (SLKMC)} in which the combination of
an automatic generation of a data base of single and multiple atom
processes during the evolution of the system, together with a
pattern recognition scheme provides a possible answer to the above
question \cite{slkmc}. Our aim here is to apply { \it SLKMC } to
examine the trends in the diffusion of small Cu islands on
Cu(111). Since in the original version of the code adatoms are
assumed to occupy only fcc hollow sites, and in the light of the
possible importance of the hcp site from the discussion above, we
have also carried out MD simulations for further insights into the
mechanisms controlling island diffusion. Indeed, the MD
simulations reveal the importance of processes involving concerted
island motion. A second set of { \it SLKMC} simulations with an
enhanced database is then performed and comparisons of the results
of the two sets of KMC simulations provide an understanding of the
factors that control the trends in the behavior, and the atomistic
processes that determine the diffusive motion of small Cu islands
on Cu(111). In particular, our results provide interesting
insights into the conditions that may lead to anomalous diffusion
coefficients for certain sizes of islands. Also, since the issue
of the relative importance of the fcc and hcp site may be system
specific, by carrying out these two sets of simulations, the
results presented here should have significance for other
surfaces.

\section{CALCULATIONAL DETAILS}

The first set of calculations are based on the recently developed
Self Learning Kinetic Monte Carlo ({ \it SLKMC})
\cite{slkmc,cghosh} technique, in which we have implemented a
pattern recognition scheme that assigns a unique label to the
environment of the diffusing atom up to several neighbors, for
efficient storage and retrieval of information on activation
energy barriers of possible processes that the system may choose
to undergo. Provisions are made for automatic calculation of the
energy barrier when a process is first identified and the result
stored in a database. These energy barriers are calculated using a
simple method which maps out the total energy of the system as the
diffusing entity moves from the initial to the aimed final site in
small steps. During the ensuing energy minimization procedure, all
atoms in the system are allowed to relax in all directions, except
for the diffusing atom whose motion is constrained along the
reaction path.  Processes involving multiple atoms can thus be
revealed naturally. Extensive comparisons of the resulting energy
barriers \cite{slkmc} with those obtained using the more
sophisticated nudged elastic band method show only minor
differences. The simpler method gives a gain of almost two orders
of magnitude in the time taken to acquire a comprehensive data
base.  For the calculation of the total energy of the system,
interatomic potentials based on the embedded atom method (EAM) are
used. The initial step in the simulation is the acquisition of the
data base. Once it has become stable, \emph{i.e.} no new processes
appear for some time (for the islands under consideration the data
base saturated after about ten million KMC steps), the system
evolves smoothly through atomistic processes of its choice and
statistics are collected for calculating quantities like the mean
square displacement center of mass of the island, correlation
functions, and the frequencies of the atomistic processes.

In the initial version of the {\it SLKMC} code, exchange processes
are not considered as the pattern recognition scheme for them is
more complex than the one implemented here for diffusion via
atomic hops.  For the diffusion of 2D islands on Cu(111), exchange
processes are also not expected to play a major role, as the
energy barrier for such processes is relatively high. Further,
they have not been identified in either experiments or in the
accompanying MD simulations performed at $500$ K.

Furthermore, we allow island atoms to occupy only fcc hollow
sites. Application \cite{altaf} of this method to the diffusion of
2D Cu islands on Cu(111) containing $19-100$ atoms has shown the
diffusion coefficient $D$ to scale with the number of atoms $N$ in
the island as $D \propto N^{-1.57}$. Periphery diffusion in which
single atoms hopped from one fcc site to another was found to be
the dominant mechanism. Several types of single and multiple atom
processes were revealed in the collected data base. However, even
for the smallest islands, the frequency of occurrence of multiple
atom processes was small and their contribution to the diffusion
process decreased further with increasing size and decreasing
temperature. In the case of islands of smaller sizes, as we have
discussed above, concerted motion might dominate the diffusion
process \cite{mar04}. These processes necessarily involve
occupation or transit through the hcp sites, although the exact
nature of such processes is not known {\it a priori}. We have thus
performed MD simulations at $500$ K to identify diffusion
processes which are not collected in the database of the present
version of the {\it SLKMC} code because of its restriction to fcc
sites. The MD simulations are performed with the same interatomic
potentials as the {\it SLKMC} simulations. As we will see the new
processes revealed in the MD simulations are beyond the ones that
could be identified by the original pattern recognition scheme of
the {\it SLKMC}. For the second set of KMC simulations we have
included all mechanisms revealed from the MD simulations in the
previously collected database of {\it SLKMC}. We have done so in
two ways, namely directly and indirectly.

In the direct method we have simply extended the pattern
recognition scheme to include occupation of the hcp hollow site.
The original $3$-shell pattern is now replaced by one with a
$9$-shell pattern which facilitates atomic jumps between the fcc
and hcp sites. All processes collected by our simple {\it SLKMC}
code and those with concerted island motion as revealed by MD were
converted into the nine shell format by hand and formed the
database of the new code which we will refer to as {\it SLKMC2}
\cite{slkmc29}. KMC simulations were performed for $1,2,3,$ and
$4$ atom islands using the {\it SLKMC2} code.  For the $5-10$ atom
islands such conversion of the pattern recognition scheme turned
out to be nontrivial.  We have thus used an indirect procedure for
the inclusion of the fcc-hcp jumps.  We have also verified that
the indirect method gives the same outcome as the direct one for
the case of the smaller islands. To incorporate the new processes
involving concerted island motion in {\it SLKMC} we have made the
following assumption. The energy barrier for an island to diffuse
collectively from the fcc to the hcp occupation
configuration is the same as the reverse process. Hence, one may
mimic two consecutive hops (fcc to hcp to fcc) by a single fcc-fcc
with half the prefactor. Using this assumption, we have used our
simple {\it SLKMC} (only fcc occupation) to study the diffusion of
all islands of sizes $2$ to $10$. For the smaller islands ($3$ and
$4$ atoms) this approximation provides the same result as obtained by
using fcc-hcp pattern recognition scheme as embedded in {\it
SLKMC2}. In Table 1 we have summarized the results of KMC
simulations for trimers and tetramers by using the direct and
indirect methods. As can be seen both yield almost the same
diffusion coefficients.

Assuming the validity of the transition state theory, the rate for
an atom to hop to a vacant site is given by $r_i = \nu_i
\exp(-\Delta E_i / k_B T)$.  Here $\Delta E_i$ is the activation
energy barrier, $k_B$ is the Boltzmann constant, and $\nu_i$ is
the attempt frequency or the so-called prefactor.  Most of the
thermodynamics of the system is hidden in the prefactor and, in
principle, it should be sensitive to the details of the atomic
environment. The prefactors for the various processes can thus be
expected to be different.  However, calculation of prefactors is
non-trivial although the recipe is well defined \cite{kurpick}.
Recent calculations of the prefactors for concerted island motion
containing $2-7$ atoms show some variation with size \cite{mar04}
but the effect is not dramatic. In principle it would be
preferable to calculate the prefactors for all the processes
present in the database.  We leave these calculations for the
future, and invoke here the often used assumption of a standard
value of $10^{12} {\rm sec}^{-1}$ for all prefactors. For further
efficiency in the KMC algorithm, we have employed the
Bortz-Kalos-Lebowitz (BKL) updating scheme \cite{bkl} which allows
one to reach macroscopic time scales of seconds or even hours for
simulations at, say, room temperature as has been shown in recent
works \cite{tapio2, tapio3}.

As for the model system, we consider a fcc(111) substrate with an
adatom island on top, as shown in Fig. 1. The gray circles are
substrate atoms which stay rigid during the simulation, whereas
the dark (colored on-line) circles are the island atoms, placed on
fcc sites which are the hollow sites having no atoms underneath
them in the layer below. A KMC simulation step begins by placing
an ad-atom island of desired size, in a randomly chosen
configuration, on the substrate. The system evolves by performing
a process of its choice, from the multitude of possible single or
multiple adatom jumps at each KMC step. We performed about $10^7$
such steps at $300$ K, $500$ K, and $700$ K. Typically, at $500$
K, $10^7$ KMC steps were equal to $10^{-3}$ sec in physical time.
The diffusion coefficient of an adatom island is calculated by $D
= \lim_{t\rightarrow \infty} \langle [R_{\rm CM}(t) - R_{\rm
CM}(0)]^2 \rangle /2 d t$, where {\it D} is the diffusion
coefficient, $R_{\rm CM(t)}$ is the position of the center of mass
of the island at time {\it t}, and $d$ is the dimensionality of
the system.

\section{RESULTS AND DISCUSSIONS}

We present first the results that are obtained from the {\it
SLKMC} method with single and multiple atom processes involving
jumps from one fcc site to another, which are automatically
accumulated and performed during the simulation.  The calculated
diffusion coefficients of the islands at $300$ K, $500$ K, and
$700$ K are summarized in Table 2. These are the numerical values
in the first entry for each size type in Table 2 and range from
$8.82 \times 10^{10}{\AA{^2/{\rm sec}}}$ for the dimer to $4.12
{\AA}^2/{\rm sec}$ for the $10$-atom island.  A log-log plot of
$D$ \emph{vs.} $N$ in Fig. 2a shows oscillations in the diffusion
coefficient with size.  This hint for magic sizes of islands
signifying reduced mobility is also seen in the Arrhenius plot of
$\ln D$ vs $1/k_BT$. The effective diffusion barriers extracted
for each island size from the Arrhenius plot (Fig. 3) also display
oscillatory behavior. As can be seen in Table 2, the $3$, $7$, and
$10$-atom islands display higher effective barriers than the
others.  The barrier for diffusion is particularly high for the
perfect hexagon ($7$-atom) island.

As we mentioned in Sec. II, MD simulations carried out at $500$ K
revealed several new concerted moves of the islands which involved
occupation of the hcp sites, too. Before discussing the details of
the atomistic processes let us examine the results for the
diffusion coefficients once these processes are included in the
database of {\it SLKMC}.  The calculated diffusion coefficients,
effective energy barriers, and the prefactors for the second set
of KMC simulations are summarized in Table 2. These values are
given in the square brackets underneath the corresponding ones
obtained when hcp-site assisted processes are not included. The
size dependence of the diffusion coefficients at three different
temperatures with the inclusion of concerted moves from MD are
also shown in Fig. 2(b) for comparison of the case already
discussed in Fig. 2(a). Further comparison of the results of the
two sets of simulations is in Fig. 3, in which the effective
diffusion energy barrier appears to scale with the island size
once the hcp-assisted concerted motion is taken into account. The
striking result is that there is no longer any oscillation in the
quantities and the $3$ and $7$ atom islands diffuse just like the
others, in proportion to their size. We now turn to an analysis of
the details of the single and multiple atom mechanisms involved in
the diffusion of the islands, one by one.

\subsection{Monomer}
For completeness we begin with a few comments on the diffusion of
an adatom on Cu(111). The primary motion for a single atom is
simply the process of hopping between the fcc and the hcp sites.
We find an activation energy barrier of $E = 29$ meV for the
process, while with a slightly different EAM potential Marinica
{\it et al.} \cite{mar04} find it to be $41$ meV. As already
mentioned, exchange processes between the adatom and the substrate
atoms are not included in our KMC simulations, neither do they
appear in the accompanying MD simulations. The effective diffusion
barrier inferred from the Arrhenius plot of the monomer
diffusivities from our KMC simulations is $E_a = 26 \pm 3$ meV,
which is consistent with the calculated energy barrier contained
in our database.  This value is also in agreement with that
obtained from MD simulations ($E_a = 31.0 \pm 0.8$ meV) by
Hynninen \emph{et al.} \cite{antti}. Experimental results report
the adatom activation energy on Cu(111) to be $E_a = 37.00 \pm 5$
meV \cite{repp}. Our results are thus in agreement with
experimental data.

\subsection{Dimer}
In the case of the motion of the dimer the {\it SLKMC} code picked
up only two mechanisms which permit jumps from fcc to fcc sites.
These are labeled  {\it Dimer A} and {\it Dimer B} in Fig. 4 (a),
and their energy barriers are $101$ meV and $15$ meV,
respectively. Results of diffusion coefficients with these two
processes at three different temperatures are in Table 2. The
effective diffusion barrier for the dimer from the Arrhenius plot
is $104$ meV. Of course, the dimer motion is actually nontrivial
since in reality both dimer atoms could also occupy the hcp sites
or they could occupy mixed sites with one atom on the fcc and the
other on the hcp site. The MD simulations actually revealed $13$
more mechanisms for the dimer diffusion which are shown in Fig. 4
(b). These illustrations show transitions between the sites
occupied by the dimer atoms. Simultaneous occupation of mixed
sites is slightly favorable because of the somewhat lower energy
(by $1 - 3$ meV), as compared to both atoms occupying the same
type of sites.

Let us have a critical look at mechanisms shown in Fig. 4 (b).
Processes describing sliding and rotational motion, $D_2, D_3,
D_4, D_6$, and $D_8$, have lower energy barriers as compared to
the others shown in Fig. 4 (b). Process $D_2$ in which both atoms
are initially on hcp sites and one jumps to the fcc site by
crossing the bridge site, has the lowest energy barrier of $5$
meV, The second low energy mechanism is $D_6$, ($9$ meV), in which
both atoms occupy fcc sites initially and one of them jumps to hcp
site by crossing the bridge site. The energy barrier for the same
mechanism from the experimental data reported by Repp \emph{et
al.} \cite{repp} is $18 \pm 3$ meV, which is larger than what we
find. Marinica \emph{et al.} \cite{mar04} find this barrier to be
$16$ meV. Process $D_4$ describes dimer atoms as initially
occupying mixed sites and finally both atoms occupy hcp sites. We
find its energy barrier to be $18$ meV while Marinica \emph{et
al.} \cite{mar04} reported it to be $E = 26$ meV. We also observed
long jump mechanisms ($ D_7, D_{10}, D_{11},$ and $D_{13}$) for
dimer diffusion in our MD simulations.

The sliding motion between fcc and hcp sites has diffusion
barriers of the same order of magnitude as that for the long jump
motion of the dimer. On the other hand, the rotational motion,
$D_3$, has a diffusion barrier ($20$ meV) closer to the value of a
single atom hopping barrier, which is $29$ meV. Finally, we
included all of these $13$ mechanisms in our {\it SLKMC} database
that had only two mechanisms ({\it Dimer A} and {\it Dimer B})
initially. As we can see from Table 2, with the inclusion of
concerted motion the effective diffusion barrier reduces to $E_a =
92$ meV, which is closer to the value of barrier representing
concerted motion of the dimer. Although a dimer performs low
energy mechanisms ($D_2, D_3, D_4, D_6,$ and $D_8$) more
frequently, the change in the center of mass position is small as
compared to the long jump mechanisms ($D_7, D_{10}, D_{11},$ and
$D_{13}$) and also concerted motion mechanism ($D_1$ and
$D_{12}$). Hence a small frequency of relatively high energy
mechanisms (long jumps) can greatly change the center of mass
position of the dimer. This is why the effective diffusion barrier
of dimer is closer to the diffusion barrier of long jumps and
concerted motion mechanisms ($D_7, D_{11}, D_{13}$).

\subsection{Trimer}
We have done a detailed study of trimer diffusion using { \it
SLKMC} simulations. There are only nine possible atom-by-atom motion
mechanisms which were identified by our {\it SLKMC} code. These
mechanisms and their corresponding energy barriers are shown in
Fig. 5 (a). With only fcc to fcc jumps the effective diffusion
barrier for the trimer is $380$ meV (see Table 2). Actually the atom-by-atom motion produces
a shape change but does not facilitate the diffusion of the trimer.
We obtained quite interesting results when we included mechanisms describing
concerted motion of trimer as shown in Fig. 5 (b). Trimer moves
from one fcc site to the neighboring hcp site by performing
concerted gliding and rotation mechanisms. The energy barrier for
concerted gliding of the trimer from 3fB to 3hA is found to be
$125$ meV, where as the reverse mechanism has a barrier of $115$
meV. The rotation of the trimer has the lowest energy barrier of
all: $38$ meV from 3hA to 3fA and  $62$ meV from 3fA to 3hA,
respectively. With the inclusion of these additional processes the
effective diffusion barrier is found to be $141$ meV.

This is a dramatic reduction from $380$ meV found earlier and the
effect is impressively represented in Fig. 2b and Fig. 3b which
shows the trimer to be relatively mobile. In Fig. 6 we plot the
distribution of the frequency of events with and without rotation
and concerted motion, represented respectively by filled and open
symbols. We find that the occurrence frequencies of added
mechanisms (concerted motion and rotation) are much higher than
the occurrence frequencies of all other nine mechanisms because
concerted motion and rotation mechanisms have low energy barriers
as compared to the mechanisms such as {\it Opening From A} and
{\it Opening From B}. Although rotation dominates, it does not
play a key role in trimer diffusion because it is not responsible
for the center of mass motion of the trimer. We expect concerted
motion to dominate diffusion, and thus we can predict that the
value of the effective diffusion barrier should be closer to the
value of the concerted motion barrier, which is indeed true here.
In Table 2, we can clearly see the difference between results
before and after including rotation and concerted motion
mechanisms in our primary database of nine processes.

\subsection{Tetramer}
In the case of the tetramer we have $28$ possible, fcc to fcc,
atom-by-atom jump processes which together with their energy
barriers are shown in Fig 7 (a). As noted in Table 2, KMC
simulations performed with these mechanisms led to an effective
diffusion barrier of $492$ meV. Three mechanisms exhibiting
concerted motion and shearing of a diamond shaped tetramer,
revealed in MD simulations, and their corresponding diffusion
barriers, are shown in Fig. 7(b). Concerted motion of a diamond
shaped tetramer takes place through sliding between the fcc and
the hcp sites, along its small and large diagonals. The ones along
the small diagonal (Fig. 7b-1) have lower energy barrier ($167$
meV for fcc to hcp and $125$ meV for hcp to fcc) than those along
the large diagonal (Fig. 7b-2). These processes have also been
discussed by Marinica \emph{et al.} \cite{mar04}. However, the
case of diamond shape tetramer diffusion through shearing
mechanism shown in Fig. 7b-3 with energy barrier $230$ meV was not
taken into account by them. When we included these three
mechanisms in our database of $28$ single atom mechanisms and
performed KMC simulation, we found significantly different values
for the diffusion coefficients. In Table 2, these values are
written in square brackets and the effective diffusion barrier for
tetramer is $E_a = 212$ meV.

\subsection{Islands containing 5 to 10 atoms}
A few examples of single atom processes collected in the database
of our KMC simulations, for the islands containing $5$ to $10$
atoms, are shown in Fig. 8  with the corresponding energy
barriers. The diffusion coefficients calculated from KMC
simulations based on these single atom mechanisms are very low as
shown in Table 2. This is particularly the case for the $7$ and $8$
atom islands whose effective diffusion barriers are consequently
the largest. This is understandable because we find that processes
such as {\it AB Corner Detachment A} and {\it AB Corner Detachment
B}, shown in Fig. 8, play a key role in the island diffusion by
contributing the most to the change in the center of mass
position. Processes such as {\it Step Edge A} and {\it Step Edge
B} occur more frequently, but they do not contribute significantly
to the motion of the center of mass of the island; rather
the atoms move around and around along the periphery of the
island. In Fig. 9 we show the concerted motion processes revealed
from MD simulations. Their energy barriers were determined from
molecular static calculations by dragging the central atom of the
island from fcc to the nearest hcp site. Other atoms in the island
followed its motion by gliding over the bridge sites. The
different shapes and geometries of these islands contribute to the
differences in the energy barriers for the processes. For example
in our MD simulations we found that the $10$ atom island can move
as a single entity from fcc to hcp sites whenever it appears into
one of the three shapes shown in Fig. 9. The energy barriers
associated with these processes are slightly different. Clearly,
the barriers of these concerted motion mechanisms (for $5$ to $10$
atom islands) are comparatively lower ($270$ meV to $590$ meV)
than the energy barriers of the single atom mechanisms {\it AB
Corner Detachment A} and {\it AB Corner Detachment B}, also
considered essential for island diffusion. After the inclusion of
new low energy concerted motion mechanisms in our database, the
high energy single atom mechanisms become less frequent in KMC
simulations and high values of diffusion coefficients and
correspondingly low values of the effective diffusion barriers
were obtained (see Table 2). The size dependent oscillations of
the diffusion coefficients and the effective diffusion barriers
also disappeared from the plots shown in Fig. 2(b) and Fig. 3,
respectively. We can thus conclude that the absence of the low
energy concerted motion mechanisms is responsible for the
oscillatory behavior of diffusion coefficients as the function of
size. Finally, our complete KMC results show that the effective
diffusion barriers increase almost monotonically with increasing
island size.

\subsection{Key mechanisms and their occurrence frequencies}

In Fig. 10 we show the normalized frequencies of all events from
the extended { \it SLKMC} data that were performed during the
simulations. Lines with open symbols in Fig. 10 show the
occurrence frequencies of the all concerted motion mechanisms, at
three different temperatures, as the function of the island size,
while those with filled symbols all single atom mechanisms. For
the dimer case, most of the single atom mechanisms have the same
effective barriers as compared to the barriers associated with
concerted motion mechanisms. Thus, the occurrence frequencies of
single and multiple atom mechanisms are almost the same for dimer
diffusion. In the case of $3$ to $7$ atom islands, concerted
motion processes are associated with significantly lower energy
barriers as compared to the single atoms, and therefore concerted
motion occurs more frequently. A six atom island has an effective
barrier for concerted motion that is closer to barriers of some
single atom mechanisms, which play a role in the motion of the
center of mass position to some extent ({\it i.e. Step Edge A} and
{\it Step Edge B} processes). Because of the close competition
between concerted motion and single atom mechanisms, we find a
narrow gap between their occurrence frequencies in the case of an
$6$ atom island. A similar, narrow gap can be seen in the case of
the $8$ atom island. In this case there is close competition
between concerted motion and the motion of the single atom going
around the periphery of the island ({\it i.e. BB Corner
Detachment} and {\it AA Corner Detachment} processes). In the case
of $9$ and $10$ atom islands the low energy single atom mechanisms
({\it i.e. Step Edge A} and {\it Step Edge B} processes) occur
more frequently, but they do not play a key role in island
diffusion. On the other hand, since the barriers of the concerted
motion mechanisms are higher ($410$ meV to $590$ meV), they occur
rarely but still play an important role in the diffusion.

\section{CONCLUSIONS}

To summarize, we have performed a systematic study of the diffusion of small Cu islands on
Cu(111), using a recently developed self learning KMC simulations in which the system is
allowed to evolve through mechanisms of its choice with the usage of a self generated database
of single and multiple atom diffusion processes. Complementary molecular dynamics simulations 
carried out for a few cases provided further details of several new mechanisms for small island diffusion which were
not automatically picked up by our {\it SLKMC} method because of the initial restriction of fcc site occupation. We found significant
changes in the size dependent variations of diffusion characteristics of the islands after including concerted motion mechanisms
which were revealed from MD simulations. We find that these small sized islands diffuse primarily through concerted motion
with a small contribution from single atom processes, even though for certain cases the frequency of single atom processes is large
because of lower activation energies. By allowing the system the possibility of evolving in time through all types
of processes of its choice, we are able to establish the relative significance of various types of atomistic processes
through considerations of the kinetics and not just the energetics and/or the thermodynamics, as is often done.  
For small Cu islands on Cu(111), we find the effective barriers for diffusion to scale with island size. We await experiments
to verify our findings.


\section{ Acknowledgements }

We thank James Evans for helpful discussions.

This work was supported by NSF-CRDF RU-P1-2600-YA-04, NSF-ERC
0085604 and NSF-ITR 0428826. T.A-N. has been supported in part by
a Center of Excellence grant from the Academy of Finland.



\newpage

FIGURE CAPTIONS: \\

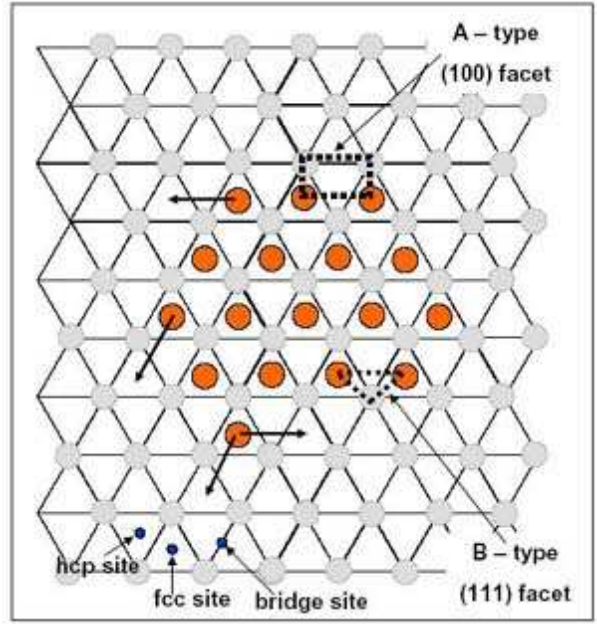
FIG. 1: Some examples of adatom diffusion and hops on fcc(111)
surface. Dark-colored atoms are active and placed at fcc sites,
whereas light-colored atoms serve as the substrate. The lower edge
of the layer containing active atoms forms a (111) micro-facet, so
it is called the B-type step edge while the upper edge of the
layer
containing active atoms forms a (100) micro-facet which is called an A-type step. \\

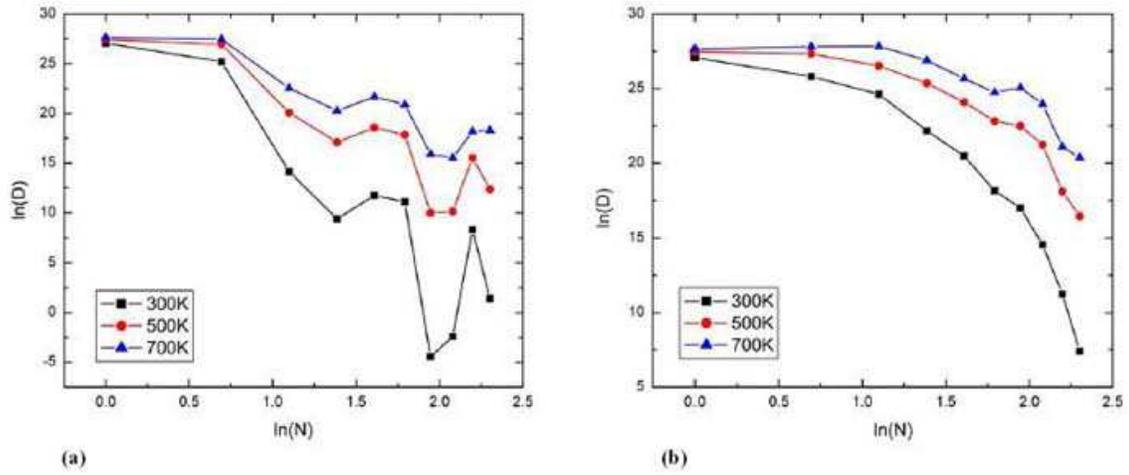
FIG. 2: Diffusion coefficients as a function of the island size;
(a) KMC results without including concerted
motion mechanisms;
(b) KMC results after including concerted motion mechanisms obtained from MD simulations. \\

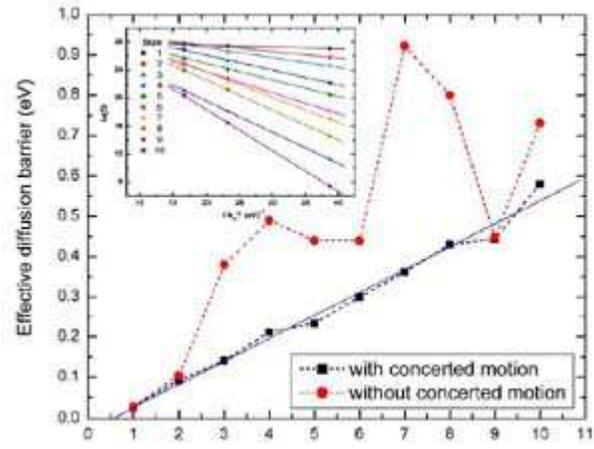
FIG. 3: Effective diffusion barriers of $1$ to $10$ atom islands
plotted as a function of island size. The dotted line with squares
represents full KMC simulation results including concerted motion,
whereas the dotted line with circles shows results of the KMC
simulation without including concerted motion mechanisms. The
inset shows Arrhenius plots of diffusion coefficients
as a function of temperature. \\

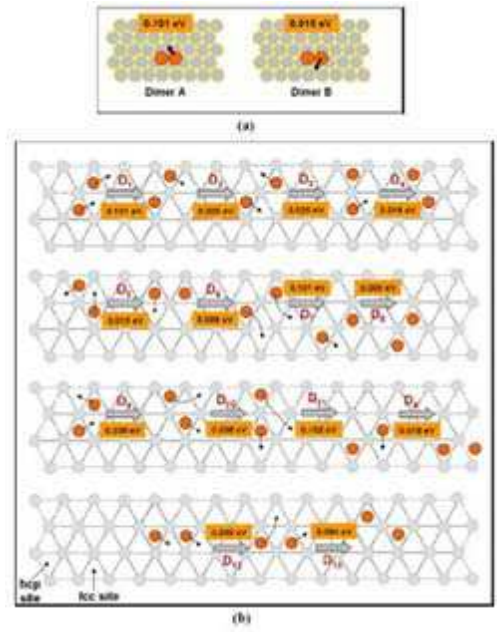
FIG. 4: (a) Illustration of two simple mechanisms for dimer
diffusion and their energy barriers, where atoms jump from fcc to
fcc sites. (b) $13$ mechanisms for dimer diffusion via fcc to hcp
sites and their energy barriers.
These mechanisms were found from MD simulations. \\

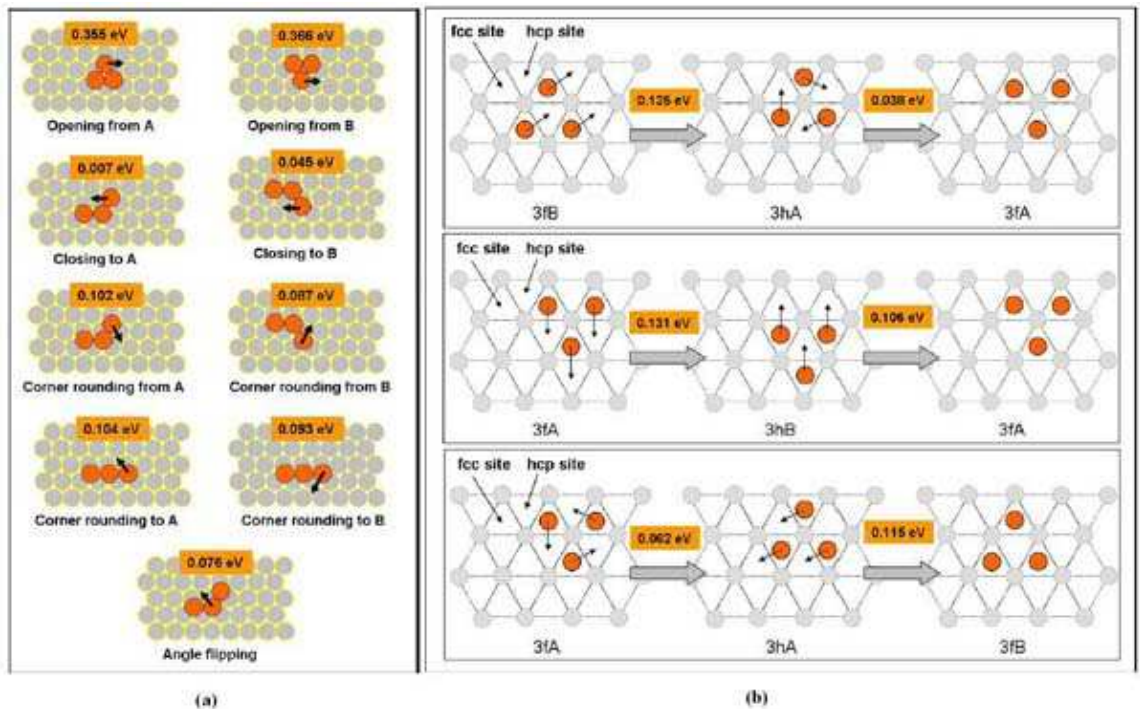
FIG. 5: (a) Nine mechanisms for trimer diffusion with their
corresponding energy barriers, where atoms are allowed to jump
from fcc to fcc sites. (b) Trimer diffusion mechanisms observed
during MD simulations. These mechanisms are conducted through a
collective motion of three atoms by rotation and gliding over the
bridge sites from fcc
to hcp sites, or vice versa. \\

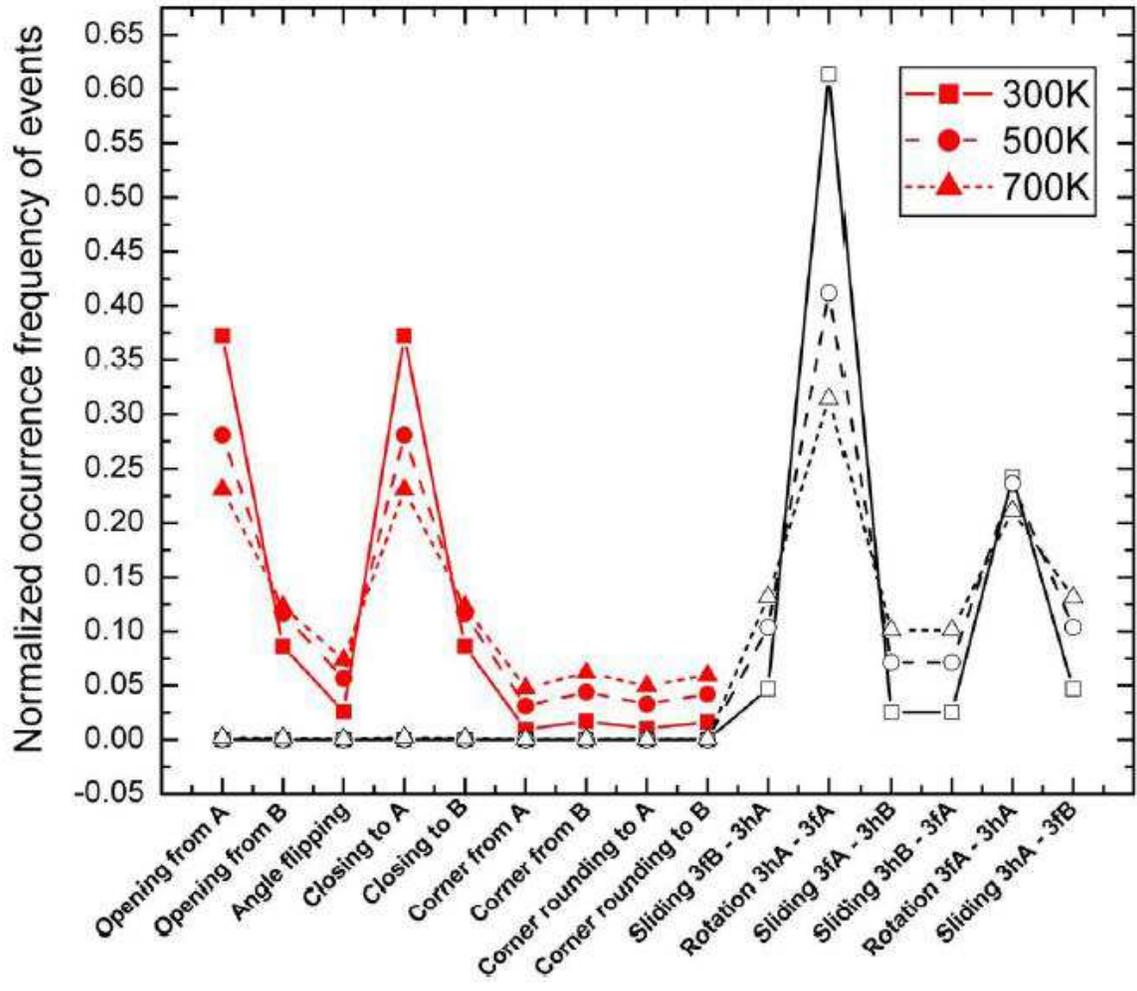
FIG. 6: Distribution of normalized frequencies of event
occurrences in the case of trimer diffusion. The lines with open
symbols show the distributions of events at different temperatures
when only single atom mechanisms were included, and lines with
filled symbols show the distributions of all events including the
collective motion of three
atoms by rotation and gliding. \\

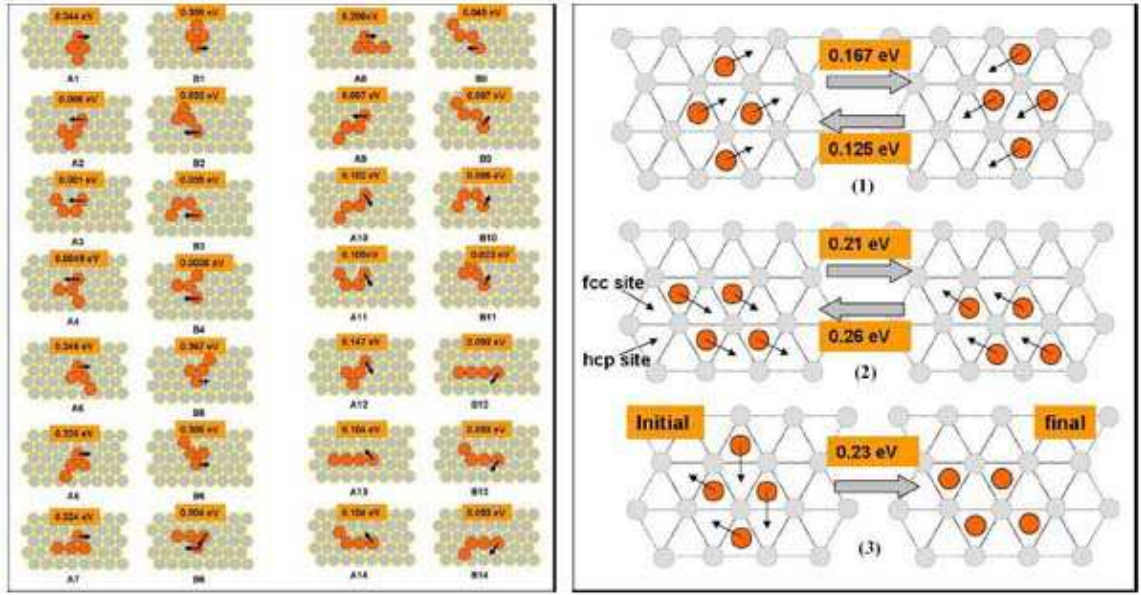
FIG. 7: (a) Illustration of $28$ mechanisms and their
corresponding energy barriers for tetramer diffusion, where jumps
are allowed from fcc to fcc sites only. (b) Tetramer diffusion
mechanisms revealed from MD simulations: (1) diagonal glide, (2)
vertical glide of $4$ atom island over the bridge sites and the
corresponding energy barriers,
and (3) shearing mechanism. \\

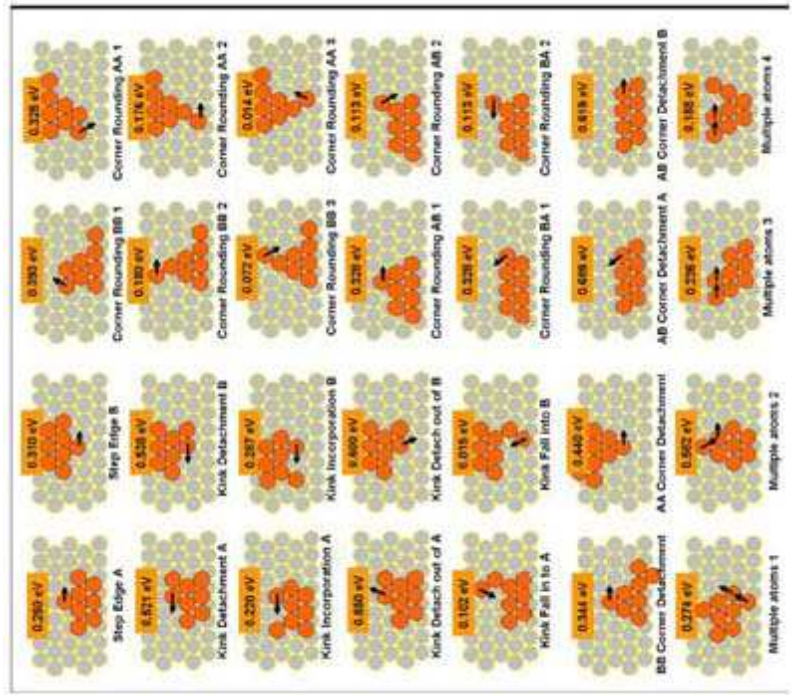
FIG. 8: A few examples of single and multiple atom mechanisms and
their corresponding energy barriers used in
our KMC simulations for islands larger than $4$ atoms.
Jumps are allowed from fcc to fcc sites only. \\

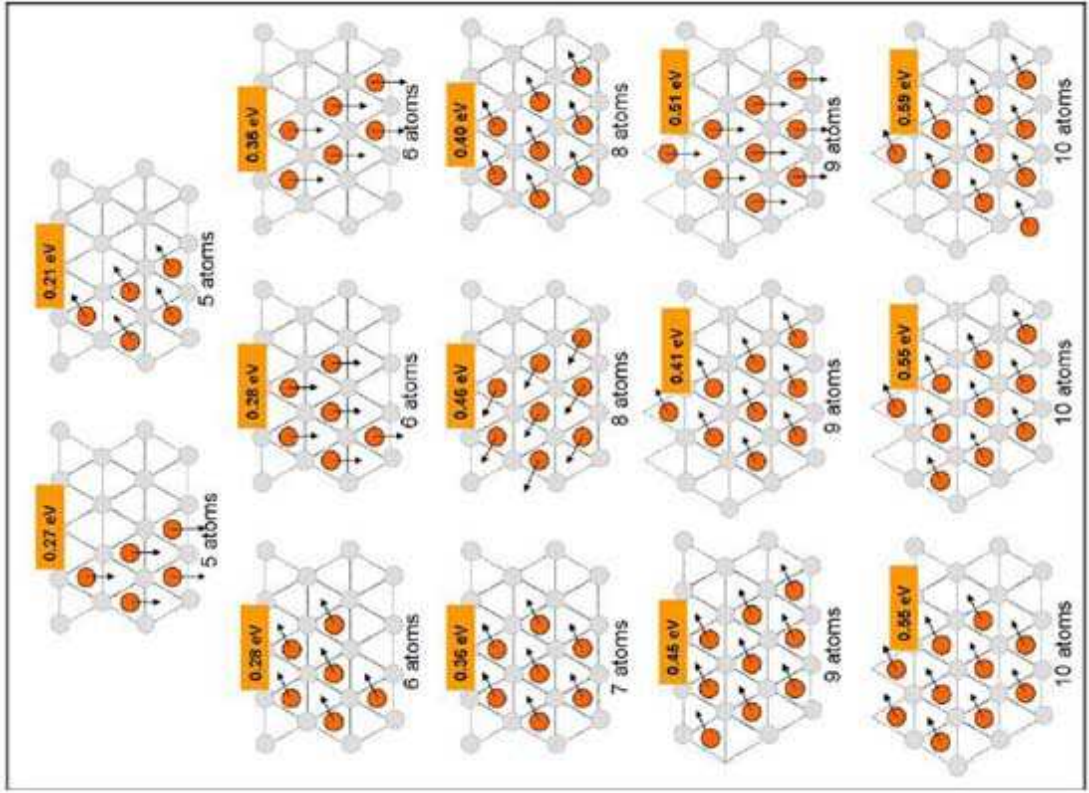
FIG. 9: Diffusion mechanisms found by performing MD simulations
for the island sizes of $5$ to $10$ atoms and
their corresponding energy barriers when they glide
over the bridge sites exhibiting collective motion of all atoms. \\

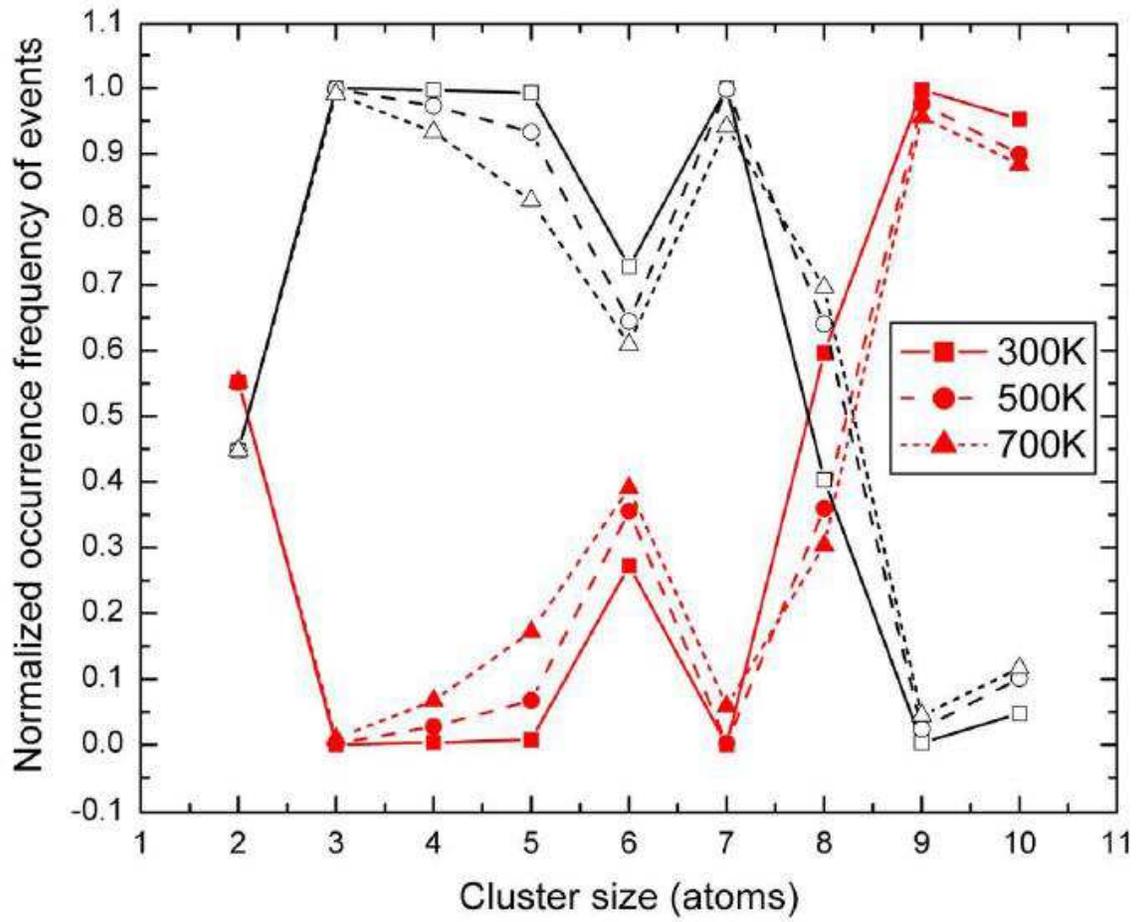
FIG. 10: Distribiution of normalized frequencies of events as a
function of the island size. The lines with open symbols represent
the frequencies of concerted motion mechanisms while the lines
with filled symbols show frequencies of mechanisms related to
single atom motion. \\

%
\newpage

TABLE 1: Diffusion coefficients of a trimer and a tetramer at
different temperatures.

\begin {tabular}{cccccc}
\hline
Island&Temperature& SLKMC &  (SLKMC+concerted motion) & [SLKMC2+concerted motion] \\
\hline
    Trimer    & 300   &$1.37\times10^{6}$ & $(2.78\times10^{10})$ &  $[4.89\times10^{10}]$ & \\
              & 500   &$5.26\times10^{8}$ & $(1.83\times10^{11})$ &  $[3.27\times10^{11}]$ & \\
              & 700   &$6.17\times10^{9}$ & $(4.55\times10^{11})$ &  $[1.22\times10^{12}]$ & \\
   Tetramer   & 300   &$1.21\times10^{4}$ & $(3.40\times10^{9}) $ &  $[4.24\times10^{9}]$  & \\
              & 500   &$2.66\times10^{7}$ & $(9.17\times10^{10})$ &  $[1.03\times10^{11}]$ & \\
              & 700   &$6.35\times10^{8}$ & $(3.38\times10^{11})$ &  $[4.80\times10^{11}]$ & \\

\hline
\end {tabular}

\newpage

TABLE 2: Diffusion coefficients of $1$ to $10$ atom islands at
different temperatures and their effective diffusion barriers with
diffusion prefactors.
\begin {tabular}{cccccc}
\hline
Island &  &Diffusion coefficient $D({\AA}^2/{\rm sec})$&  &Effective & Diffusion \\
size    &   & KMC  [KMC+Concerted motion] &      & barrier & prefactor  \\
(atoms)     & 300 K &  500 K & 700 K &           $E_a$ (eV)      &   $D_0({\AA}^2/{\rm sec})$  \\
\hline
    1     & $5.70\times10^{11}$ &  $8.50\times10^{11}$ & $1.02\times10^{12}$ & 0.026 & $1.56\times10^{12}$ \\
    2     & $8.82\times10^{10}$ &  $5.07\times10^{11}$ & $8.50\times10^{11}$ & 0.104 & $5.14\times10^{12}$ \\
          & $[1.62\times10^{11}]$ &  $[7.39\times10^{11}]$ & $[1.21\times10^{12}]$ & [0.092] & $[5.86\times10^{12}]$ \\
    3     & $1.37\times10^{6}$ &  $5.26\times10^{8}$ & $6.17\times10^{9}$ & 0.380 & $3.52\times10^{12}$ \\
          & $[4.89\times10^{10}]$ &  $[3.27\times10^{11}]$ & $[1.22\times10^{12}]$ & [0.141] & $[1.06\times10^{13}]$ \\
    4     & $1.21\times10^{4}$ &  $2.66\times10^{7}$ & $6.35\times10^{8}$ & 0.492 & $2.31\times10^{12}$ \\
          & $[4.24\times10^{9}]$ &  $[1.03\times10^{11}]$ & $[4.80\times10^{11}]$ & [0.212] & $[1.53\times10^{13}]$ \\
    5     & $1.25\times10^{5}$ &  $1.16\times10^{8}$ & $2.60\times10^{9}$ & 0.440 & $4.13\times10^{12}$ \\
          & $[7.81\times10^{8}]$ &  $[2.87\times10^{10}]$ & $[1.40\times10^{11}]$ & [0.234] & $[6.73\times10^{12}]$ \\
    6     & $6.66\times10^{4}$ &  $5.58\times10^{7}$ & $1.19\times10^{9}$ & 0.440 & $1.69\times10^{12}$ \\
          & $[7.57\times10^{7}]$ &  $[8.15\times10^{9}]$ & $[5.60\times10^{10}]$ & [0.300] & $[8.22\times10^{12}]$ \\
    7     & $1.18\times10^{-2}$ &  $2.18\times10^{4}$ & $8.00\times10^{6}$ & 0.922 & $3.80\times10^{13}$ \\
          & $[2.40\times10^{7}]$ &  $[5.80\times10^{9}]$ & $[7.60\times10^{10}]$ & [0.362] & $[2.90\times10^{13}]$ \\
    8     & $9.00\times10^{-2}$ &  $2.53\times10^{4}$ & $5.49\times10^{6}$ & 0.800 & $3.70\times10^{12}$ \\
          & $[2.10\times10^{6}]$ &  $[1.65\times10^{9}]$ & $[2.59\times10^{10}]$ & [0.430] & $[3.61\times10^{13}]$ \\
    9     & $4.18\times10^{3}$ &  $5.50\times10^{6}$ & $8.00\times10^{7}$ & 0.448 & $1.55\times10^{11}$ \\
          & $[7.72\times10^{4}]$ &  $[7.20\times10^{7}]$ & $[1.45\times10^{9}]$ & [0.444] & $[2.24\times10^{12}]$ \\
   10     & 4.12                 &  $2.33\times10^{5}$ & $8.82\times10^{7}$ & 0.731 & $7.24\times10^{12}$ \\
          & $[1.65\times10^{3}]$ &  $[1.37\times10^{7}]$ & $[7.02\times10^{8}]$ & [0.580] & $[1.06\times10^{13}]$ \\

\hline
\end {tabular}


\end{document}